\documentclass[a4paper]{jpconf}
\usepackage{graphicx}
\begin{document}

\title{Evidence for magnetic clusters in Ni$_{1-x}$V$_{x}$ close to the quantum critical concentration}

\author{R Wang$^1$, S Ubaid-Kassis$^{1,}$\footnote[2]{Present address: Dominican University of California, San Rafael, CA 94901}, A Schroeder$^1$, P J Baker$^3$, F L Pratt$^3$, S J Blundell$^4$, T Lancaster$^{4,}$\footnote[5]{Present address: Durham University, Centre for Materials Physics, South Road, Durham DH1 3LE, UK}, I Franke$^4$, J S M{\"o}ller$^{4,}$\footnote[6]{Present address: Neutron Scattering and Magnetism, Laboratory for Solid State Physics, ETH Z\"{u}rich, CH-8093 Z\"{u}rich, CH}, T Vojta$^7$  }

\address{$^1$ Department of Physics, Kent State University, Kent OH 44242, USA}
\address{$^3$ ISIS Facility, STFC Rutherford Appleton Laboratory, Harwell Oxford, OX11 0QX, UK}
\address{$^4$ Department of Physics, Clarendon Laboratory, Parks Road, Oxford, OX1 3PU, UK  }
\address{$^7$ Department of Physics, Missouri University of Science \& Technology, Rolla, MO 65409, USA }
\ead{aschroe2@kent.edu}

\begin{abstract}
The d-metal alloy Ni$_{1-x}$V$_{x}$ undergoes a quantum phase transition from a ferromagnetic ground state to
a paramagnetic ground state as the vanadium concentration $x$ is increased. We present magnetization,
ac-susceptibility and muon-spin relaxation data at several vanadium concentrations near the
critical concentration $x_c \approx11.6~\%$ at which the onset of ferromagnetic order is suppressed to zero temperature.
Below $x_c$, the muon data reveal a broad magnetic field distribution indicative of
long-range ordered ferromagnetic state with spatial disorder.   We show evidence of magnetic clusters  in the
ferromagnetic phase and close to the phase boundary in this disordered itinerant system as an important generic
ingredient of a disordered quantum phase transition.
In contrast, the temperature dependence of the magnetic susceptibility above $x_c$ is best described in terms
of a magnetic quantum Griffiths phase with a power-law distribution of fluctuation rates of dynamic magnetic clusters.
At the lowest temperatures, the onset of a short-range ordered cluster-glass phase is recognized by an increase in the muon depolarization
in transverse fields and maxima in ac-susceptibility.
\end{abstract}

\section{Introduction}
Ferromagnetism (FM) in the simple metal Ni can be suppressed by substituting Ni by another d-element like V, leading to a quantum phase transition (QPT) into a paramagnetic (PM) phase \cite{Ubaid10}. Ni$_{1-x}$V$_x$  is of special interest because it does not remain a clean, homogeneous system, but becomes strongly disordered when driven through the QPT. The disorder is controlled by the V-concentration: each V atom acts as a local impurity and forms the center of a large nonmagnetic region, i.e., a hole in the magnetization density of  the original ferromagnetic Ni. Due to the strong disorder, an unconventional QPT scenario is expected. Bulk magnetization measurements in Ni$_{1-x}$V$_x$ \cite{Ubaid10} indeed confirmed the signatures of a disordered itinerant QPT with Heisenberg symmetry \cite{vojta05,vojta09,vojtaLT}. These measurements also revealed a quantum Griffiths phase (GP) on the paramagnetic side of the transition, featuring a spectrum of fluctuating magnetic clusters. It is characterized by power laws with non-universal exponents $\alpha(x)$ in magnetization-field curves \cite{Ubaid10}. The $x$-dependence of the exponent leads to an estimate of a quantum critical concentration $x_c=11.6~\%$.
First zero-field $\mu$SR data \cite{muonpaper} revealed a broad field distribution within the FM state which scales with the mean magnetic moment for $x>x_c/2$.
This confirms a static inhomogeneous magnetization density in this compound.
As is the case for other disordered itinerant ferromagnets such as CePd$_{1-x}$Rh$_x$ \cite{wester,adroja}, the phase diagram of Ni$_{1-x}$V$_x$ is actually even more complex: At very low temperatures and $x\ge x_c$, the GP of fluctuating independent clusters undergoes a freezing transition into a cluster glass.

In the present paper, we concentrate on the low-temperature phase boundaries of the ferromagnetic and cluster glass phases. We characterize the disorder in this system and probe how the resulting magnetic inhomogeneities change from the FM phase, through the cluster glass phase, and into the paramagnet.  We pay particular attention to the formation of magnetic clusters, using several complementary methods such as bulk magnetization, ac-susceptibility and muon spin rotation ($\mu$SR) \cite{musr,amato} measurements.

\section{Experimental methods}
Polycrystalline Ni$_{1-x}$V$_x$ samples with $x=0-12.3~\%$ were grown as spherical pellets as before \cite{Ubaid10}. Magnetization and ac-susceptibility measurements were carried out in SQUID magnetometer and in pick up coil in dilution refrigerator and He$^4$ cryostat as described in \cite{Ubaid10}. The $\mu$SR data in zero (ZF) and transverse field (TF) were collected at Dolly at PSI and MuSR at ISIS using 7-30 pellets of each composition wrapped in silver foil.

\section{Magnetic susceptibility}
In our original study of the QPT in Ni$_{1-x}$V$_x$ \cite{Ubaid10}, clear signs of magnetization inhomogeneities were only observed for $x>11~\%$, where the Arrott plot (AP) description failed. Here, we return to this question, paying particular attention to small fields.

As expected for weak itinerant FMs, our samples show good AP behavior (parallel isotherms in plots of $M^2$ vs.\ $H(M)/M$) for $x$ up to $11~\%$ but only at higher fields, $H>5$~kOe.
For example, Fig. 1(a) shows that the AP fit deviates from the low-temperature $M(H)$ data for x=11~\% for fields $H<5$~kOe, leading to an overestimate of the spontaneous magnetization $M_s$. Usually some deviations at lower fields are attributed to domain orientation or other non-intrinsic effects and are disregarded. However, Ni$_{1-x}$V$_x$ is a very soft magnet without mechanical treatment and does not show any visible hysteresis.  An alternative description that better describes the low-field data is given by a power law with an offset, $M=M_0+aH^{\alpha}$. The exponent $\alpha(x)$ is nonuniversal, analogous to the Griffiths exponents ($\lambda=\alpha=1-\gamma$  identified for  $x>x_c$ \cite{Ubaid10} e.g. $\alpha(x=11~\%)\approx \alpha(x=12.3~\%)\approx 0.5$). It can be described by  $\alpha(x) \sim |x-x_c|^{\nu\psi}$ with  $x_c=11.6~\%$.  $M_0$ seems a better estimate of the spontaneous magnetization than the AP extrapolation $M_s$ in this inhomogeneous compound. The difference $(M_s-M_0)$ stems from magnetic clusters that are decoupled from the bulk ferromagnet and thus only align gradually in a small field.
The cluster fraction $P_{cluster}=(M_s-M_0)/M_s$ is negligible for small $x$, $P_{cluster}(x=9~\%)<1~\%$, but increases rapidly  towards $x_c$ (from $P_{cluster}(x=11.0~\%)=16~\%$  to $P_{cluster}(x=11.4~\%)=80~\%$). All these observations are in qualitative agreement with a quantum Griffiths phase on the ferromagnetic side of the QPT. A detailed quantitative analysis of this phase will be given elsewhere \cite{GPFM}.
Signs of magnetic inhomogeneities due to holes and clusters for concentrations $x\leq 11~\%$ were also seen in recent muon spin relaxation data that show distributions of local fields as soon as $x\ge4~\%$ \cite{muonpaper}.

Estimates of the critical temperature $T_c$ from the magnetization data also rely on extrapolations to zero field.
Figure 1(b) shows that plots of $dM/dH(T;H)$ vs.\ $T$ feature a maximum at $T_{max}(H)$ which is $H$-dependent \cite{as10}.
Extrapolating the low-field ($H<5$~kOe) values of $T_{max}(H)$ to zero field yields an estimate $T_{c,0}$ which differs
from the high field extrapolation $T_{c,hi}$ (or the corresponding AP extrapolation), see Fig.\ 1(c).
The reason is that a higher field helps to align smaller decoupled clusters already at higher $T$.
The change of slope in $M(H)$ and $T_{max}(H)$ at smaller fields and the resulting difference between $T_{c,hi}$ and $T_{c,0}$ can be viewed as characteristics of an inhomogeneously ordered system exhibiting magnetic clusters within the long-range
ordered FM state and above $T_c$.

\begin{figure}[t]
\begin{minipage}{17pc}
\includegraphics[width=17pc]{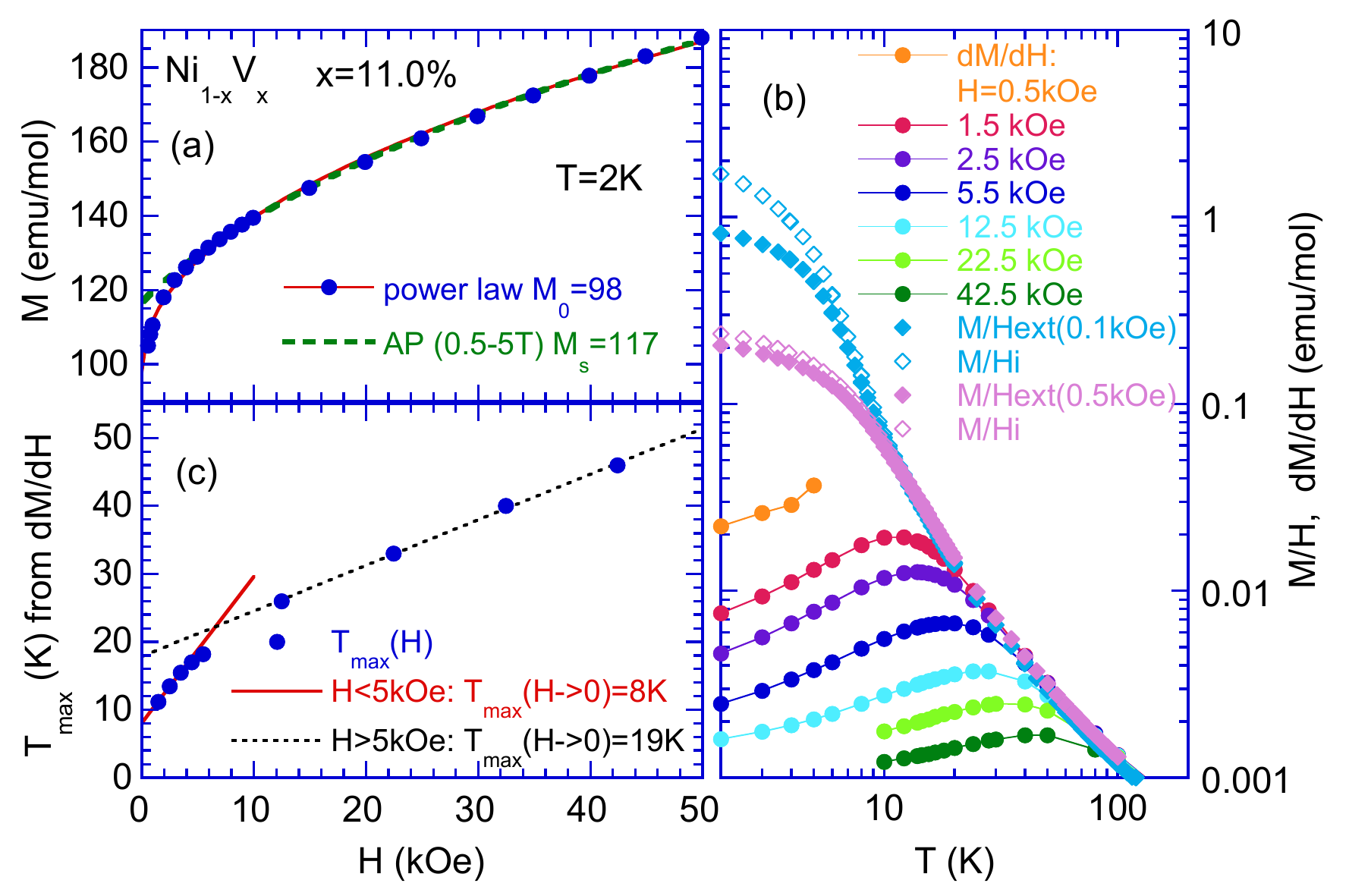}
\caption{\label{Fig1} dc-magnetization $M(H,T)$ of Ni$_{1-x}$V$_x$ with $x=11.0~\%$ vs. magnetic field $H$ (a) and vs. temperature $T$ (b). Panel (b) also shows dM/dH exhibiting a maximum at $T_{max}$(H) at different $H$ displayed in (c).}
\end{minipage}\hspace{1.5pc}%
\begin{minipage}{19.pc}
\includegraphics[width=18pc]{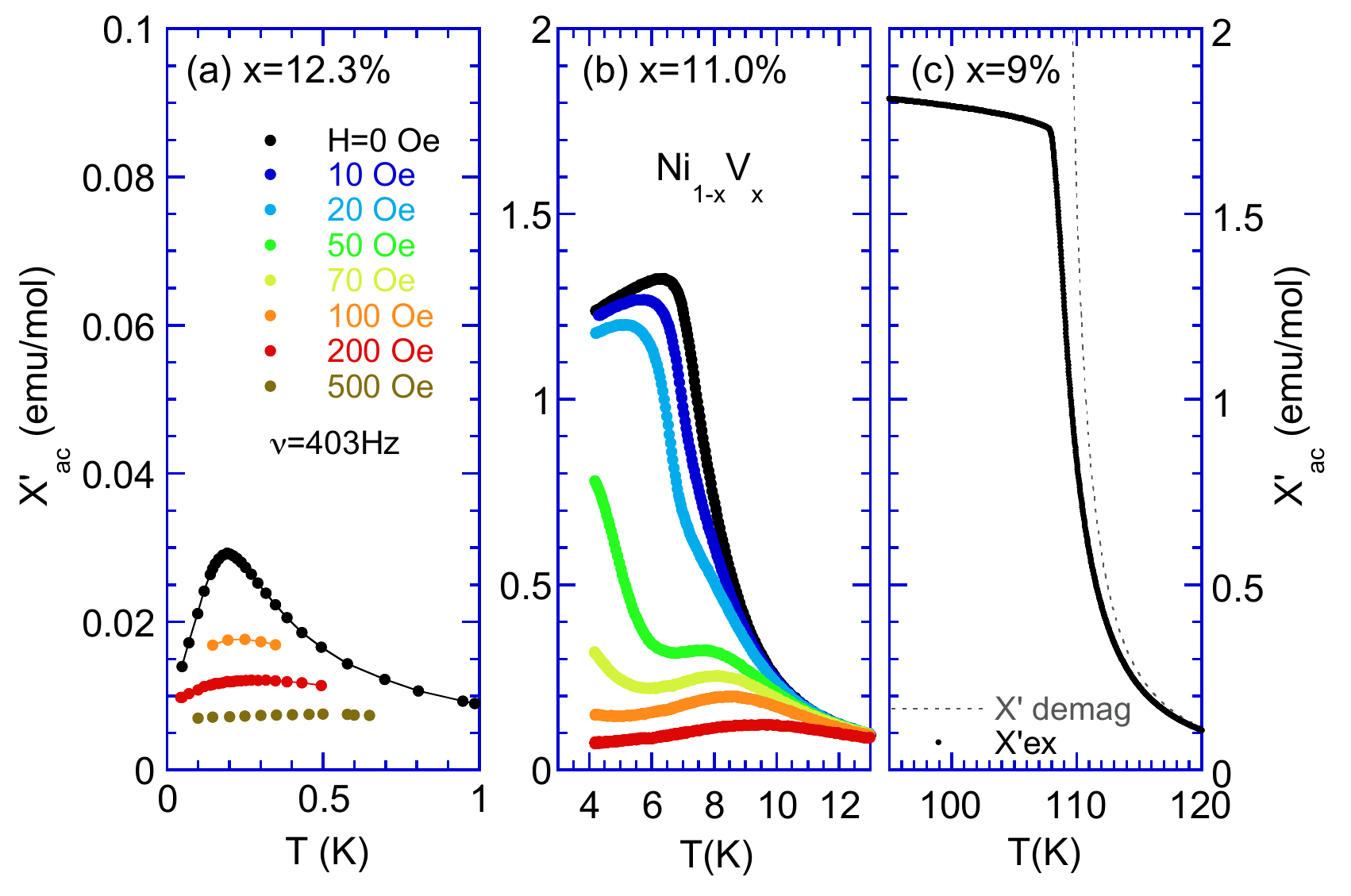}
\caption{\label{Fig2} ac-susceptibility $\chi'_{ac}$ of  Ni$_{1-x}$V$_x$ for $x=12.3~\%$ \cite{Ubaid10} (a), $11.0~\%$ (b), and 9.0~\% (c) vs temperature $T$. Shown is $\chi$ (not demagnetized) in $H_{ac} \approx 0.1$~Oe with frequency $\nu$=403~Hz. The label marks the external dc-field $H$. (absolute scale calibrated by dc-M within $30~\%$)}
\end{minipage}
\end{figure}

Figure 2 presents the ac susceptibility $\chi_{ac}$ in zero and very small external dc-fields for V-concentrations $x=9~\%$ (on the FM side), $11~\%$, and $12.3~\%$ (on the PM side). The FM transition is recognized by a sharp increase in $\chi(T)$. The figure shows the original in phase signal $\chi'=\chi=dM(H_{ext})/dH_{ac,ext}$ of spherical samples in $H_{ac}\approx0.1$~Oe and in $H_{dc}=0 - 500$~Oe. The demagnetized $\chi_{int}=dM/dH_{int}$ is significantly increased approaching the FM regime when $H_{int}$ gets reduced, while $\chi$ is limited. We take the cusp as a measure of $T_c$ for $x=9~\%$. The response for $x=12.3~\%$ is much smaller and shows no demagnetization effects. $\chi(T)$ displays a maximum at $T_{max}=T_f$ which increases with frequency $\nu$ of the ac-field as $(\delta T/T_f) / \log\nu = (0.095\pm0.005)/\textrm{decade}$ \cite{Ubaid10}. This is a signature of the onset of a cluster glass. The sample with $x=11~\%$ clearly displays a FM response, a sharp increase of $\chi$ towards low $T$,  similar to the $9~\%$ sample. However, the cusp is less pronounced here, and a broad maximum is visible at $T_{max}$ indicating $T_{c,0}$. This also hints at some antiferromagnetic couplings between the clusters. A frequency dependence of $\chi$ could not be resolved here as $\delta (T/T_c) / \log\nu < 0.005/\textrm{decade}$.  Since $H_{int}$ is not constant, a quantitative analysis is not possible, but it is noticeable that small external dc-fields change and suppress the response, indicating a spectrum of different clusters.

\begin{figure}[t]
\begin{minipage}{14pc}
\includegraphics[width=14pc]{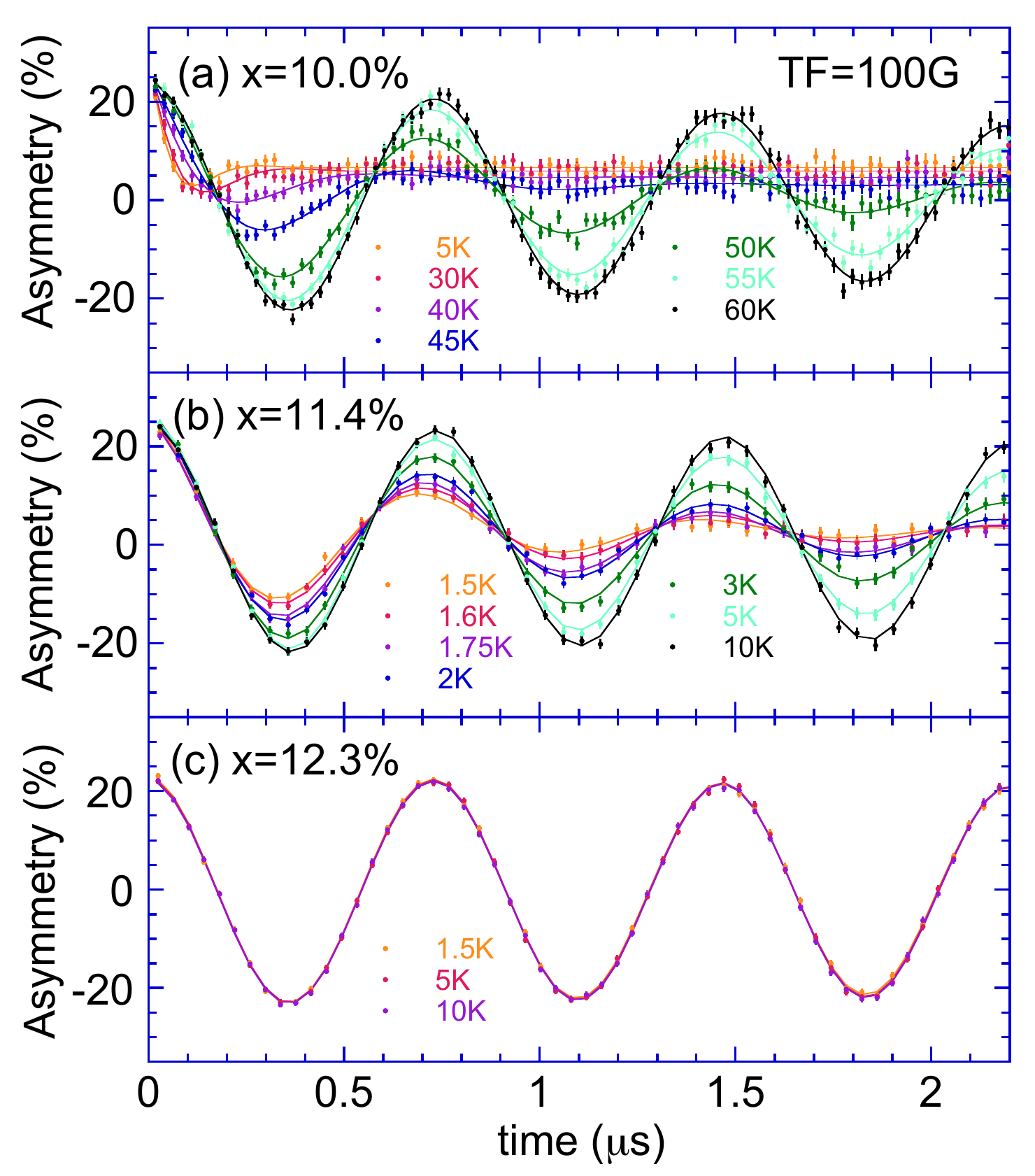}
\caption{\label{Fig3}Muon asymmetry vs. time  for different $x$ in a TF of 100~G for different temperatures $T$. Lines are  fits to Eq.(1) }
\end{minipage}\hspace{2pc}%
\begin{minipage}{21.5pc}
\includegraphics[width=21.5pc]{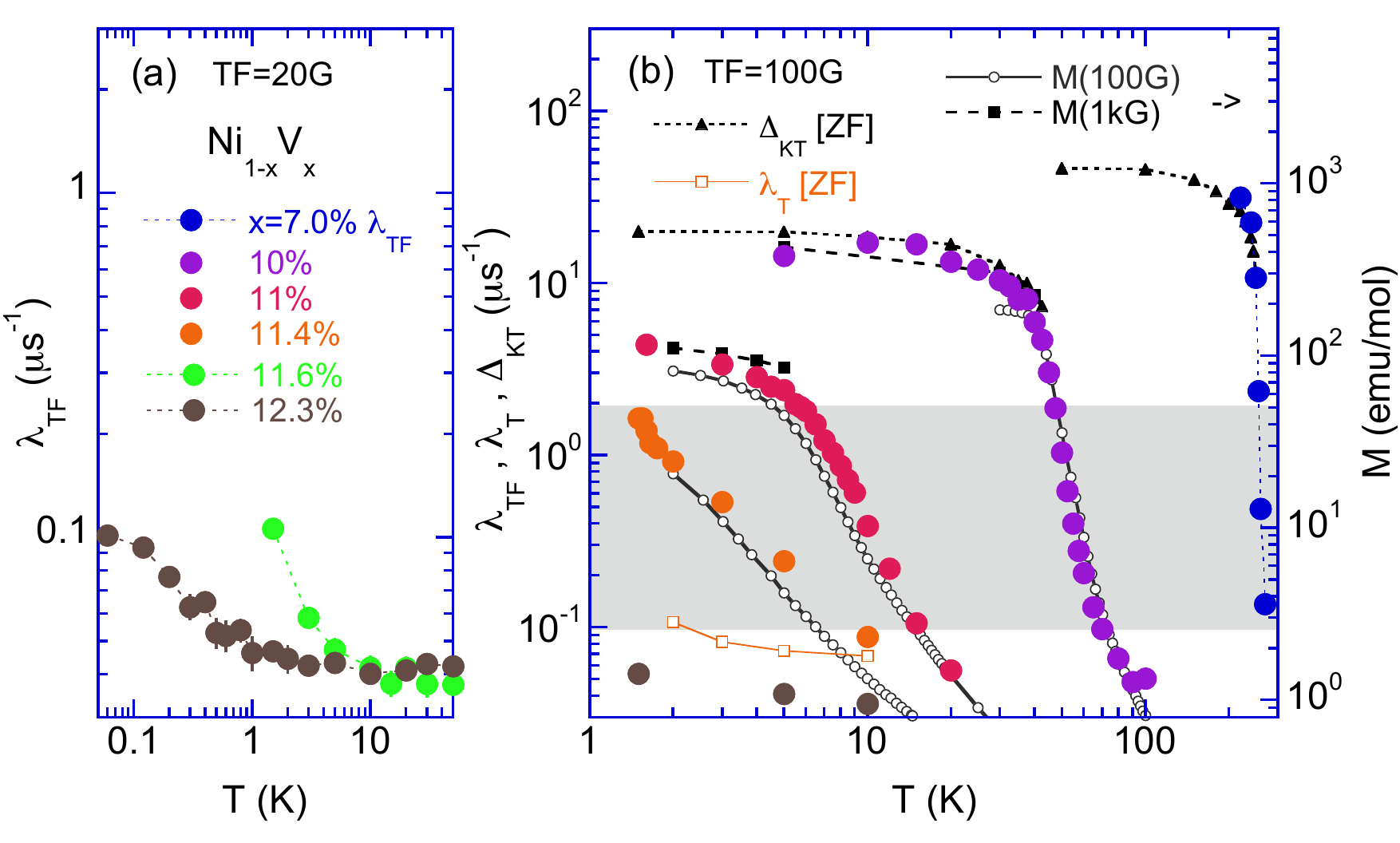}
\caption{\label{Fig4} Muon depolarization rate $\lambda_{TF}$ vs. $T$ in TF=20~G for $x>11.4\%$ (a) and  TF=100~G for $x=7.0 -12.3\%$ (b). Also shown are $\Delta_{KT}$ and $\lambda_{T}$ from ZF muon data \cite{muonpaper} and magnetization data $M(B=100$~G) and $M(B=1$~kG). The grey regime marks a PM cluster regime above $\lambda_{TF}\approx0.1~\mu$s$^{-1}$; above $\lambda_{TF}\approx2~\mu$s$^{-1}$ FM ordering occurs.}
\end{minipage}
\end{figure}

\section{Muon spin rotation}

Results of zero-field (ZF) $\mu$SR experiments were reported in Ref.\ \cite{muonpaper}. They revealed a broad magnetic field distribution within the FM state. Here we focus on $\mu$SR  data in transverse fields (TF) as a microscopic probe of the different magnetic phases and transitions in Ni$_{1-x}$V$_x$.

Figure 3 shows an overview of the muon asymmetry of three samples with $x=10$~\%, $11$~\% and $12.3$~\%  in a transverse field of 100~G. At higher temperatures (in the PM regime), nearly undisturbed muon precession is observed. Upon lowering $T$, the oscillations appear more ``damped" because of dephasing by internal magnetic fields. Note that the muons see a distribution of local fields due to the disorder in the system.
The width of the local field distribution $\Delta/\gamma_{\mu}$ (with $\gamma_{\mu}=2\pi\times 135.5$~MHz/T) can be estimated from the ZF results \cite{muonpaper}. For $x=7-10~\%$, the ZF data are well described by the Kubo-Toyabe form \cite{hayano} implying a Gaussian field distribution. Its width is generally  higher than our applied transverse field of $8.5~\mu$s$^{-1}/\gamma_{\mu}$ (for example, $\Delta=30~\mu$s$^{-1}$ for $x=10~\%$).

For $x=10~\%$, the response shows a single dip at low $T$ below $T_c$  similar to the ZF response.
In contrast, the muon precession for $x=12.3~\%$ remains nearly undamped for all $T$ down to 1.5~K (see Fig. 3(c)).
For $x=11.4~\%$, the muon precession decay rate increases with decreasing $T$ but precession remains evident at the lowest $T$ (as shown in Fig.3 (b)), indicating that long-range FM order has not formed.

In order to analyze the TF asymmetry data, we use a simple exponentially damped oscillation~(1) and a constant term~(2) describing the signal from the sample, and an undamped oscillation~(BG) describing the background from the silver mount.
\begin{equation}
A_{osci} (t; \omega, \lambda _{TF}) = A_1 \exp(-\lambda_{TF} t) \cos(\omega t + \phi) + A_2 + A_{BG}(t;\omega_{BG},\phi).
\end{equation}
$\omega$ is the observed angular frequency (in the PM regime, it is essentially given by the external transverse field, $\omega=\gamma_{\mu} H$). $A_1, A_2$ are the amplitudes of the dominating transverse response and of a weak longitudinal response (which evolves in the FM state due to longitudinal fields). $A_{BG}(t)$ has a small amplitude and the same phase, $\phi$, as (1) determined by the initial orientation of the muons relative to the detectors.
The smallest  $\lambda_{TF} \approx 0.04~\mu$s$^{-1}$ detected might be the depolarization rate due to diluted nuclear V-spins. We used a single exponential form in Eq. (1)  for all $T$. A Gaussian form did not improve the fit results at lower $T$.

Fig. 4 shows the relaxation rate $\lambda_{TF}$ extracted from Eq.\ (1) versus $T$ for several Ni$_{1-x}$V$_{x}$ samples in transverse fields of 20~G (a) and 100~G (b). With decreasing temperature, $\lambda_{TF}(T)$ increases steeply marking the onset of magnetic order. The FM samples with $x=7~\%$ and $10$~\% exhibit the sharpest increase and the highest saturation value $\lambda_{max}$ reflecting strong internal magnetic fields.  In the $x=11~\%$ sample, $\lambda_{TF}$ increases more slowly before leveling off at low $T$; while $\lambda_{TF}$ of the $11.4~\%$ sample just keeps rising without saturation down to 1.5~K. Beyond 11.4~\%, $\lambda_{TF}$ remains much smaller. These data are taken in 20~G (see Fig.4(a)). A small increase at very low $T$ likely stems from weak internal fields associated with cluster glass order.

The large relaxation rates in the FM state are mainly due to variations of the local fields because they exceed the external TF.
We see in Fig. 4(b) that $\lambda_{max}$ is close to the width $\Delta_{KT}$ of the Gaussian local field distribution from the Kubo-Toyabe fit of the ZF data for $x\leq10~\%$.  In Ref.\ \cite{muonpaper}, we showed that $\Delta_{KT}$ can be related quantitatively to the mean magnetic moment for $7~\%<x<10~\%$ by using the calibration that in pure Ni, a moment of  $m_s=0.6~\mu_B$ leads to a muon angular precession frequency of $\omega=127~$rad/$\mu$s.
Figure 4(b) compares the bulk magnetization $M$ with the relaxation rate $\lambda_{TF}$ using the same calibration. The high-field magnetization $M$(1~kG)   (taken as an estimate of the low-$T$ spontaneous magnetization) agrees well with $\lambda_{max}$ for $x=10~\%$ and $11~\%$. We see that recording $\lambda_{max}$ gives an estimate of the spontaneous moment in this disordered FM.

In the PM state, $\lambda_{TF}$ follows the $T$-dependence of the bulk magnetization in the same external field of 100~G.
The main contribution to $\lambda_{TF}$ comes from inhomogeneous field broadening:
The local field induced by the applied field is not homogeneous and can therefore dephase the muon precession in the applied transverse field. The induced local field variations are proportional to the mean inner field which is directly related to the magnetization. Thus magnetization $M$ and relation rate $\lambda_{TF}$ are expected to feature the same $T$-dependence.
Our data thus show  that magnetic clusters are present already above $T_c$. Similar behavior is observed in the disordered FM Pd-Mn($2~\%$) \cite{PdMn} which becomes a spin glass by tuning the impurity concentration: The muon relaxation rate in transverse fields is proportional to the magnetization confirming randomly placed spins. In contrast,  homogeneous itinerant magnets of stoichiometric compounds show only a very weak $\lambda_{TF}$ well above the FM transition $T_c$ \cite{REAl2}.
Spin correlation and dynamics are better probed in zero and longitudinal fields \cite{TFMnSi} undisturbed by dephasing. We see here that the depolarization rate $\lambda_T$ in ZF is much smaller than in TF in this PM regime and only increases close to $T_c$ when the clusters start freezing.

To extract values of the critical temperature from the muon data, we identify the onset of the FM with the point at which $\lambda_{TF}$ exceeds $2~\mu$s$^{-1}$. This value is of the order of the transverse field and smaller than $\lambda_{max}$ of the FM samples.
The resulting critical temperatures  match the  estimates $T_{c,0}(x)$ from the zero-field susceptibility (see Fig. 5).
The first upturns of $\lambda_{TF}$ due to magnetic clusters are noticed roughly at $\lambda_{TF}=0.1~\mu$s$^{-1} $, marking a higher $T_{c,hi}$[TF] (see grey regime in Fig.4 (b)). It matches the $T_{c,hi}$ estimates from the AP extrapolations. The onset of FM order at $T_{c,0}(x)$ decreases more rapidly with $x$ than $T_{c,hi}(x)$ (e.g. from $T_{c,0}/T_{c,hi}(x=10~\%)=70~\%$ to  $T_{c,0}/T_{c,hi}(x=11~\%)=40~\%$), indicating that the fraction of spins contributing to long range order decreases with dilution $x$.
To extract values of the cluster-glass temperature $T_f$, we take $\lambda=0.08~\mu$s$^{-1}$ as a marker of the freezing transition. This yields values that match  $T_f [\chi]$ from the maximum in $\chi(T)$\cite{Ubaid10,as10}.

\begin{figure}[t]
\includegraphics[width=17.5pc]{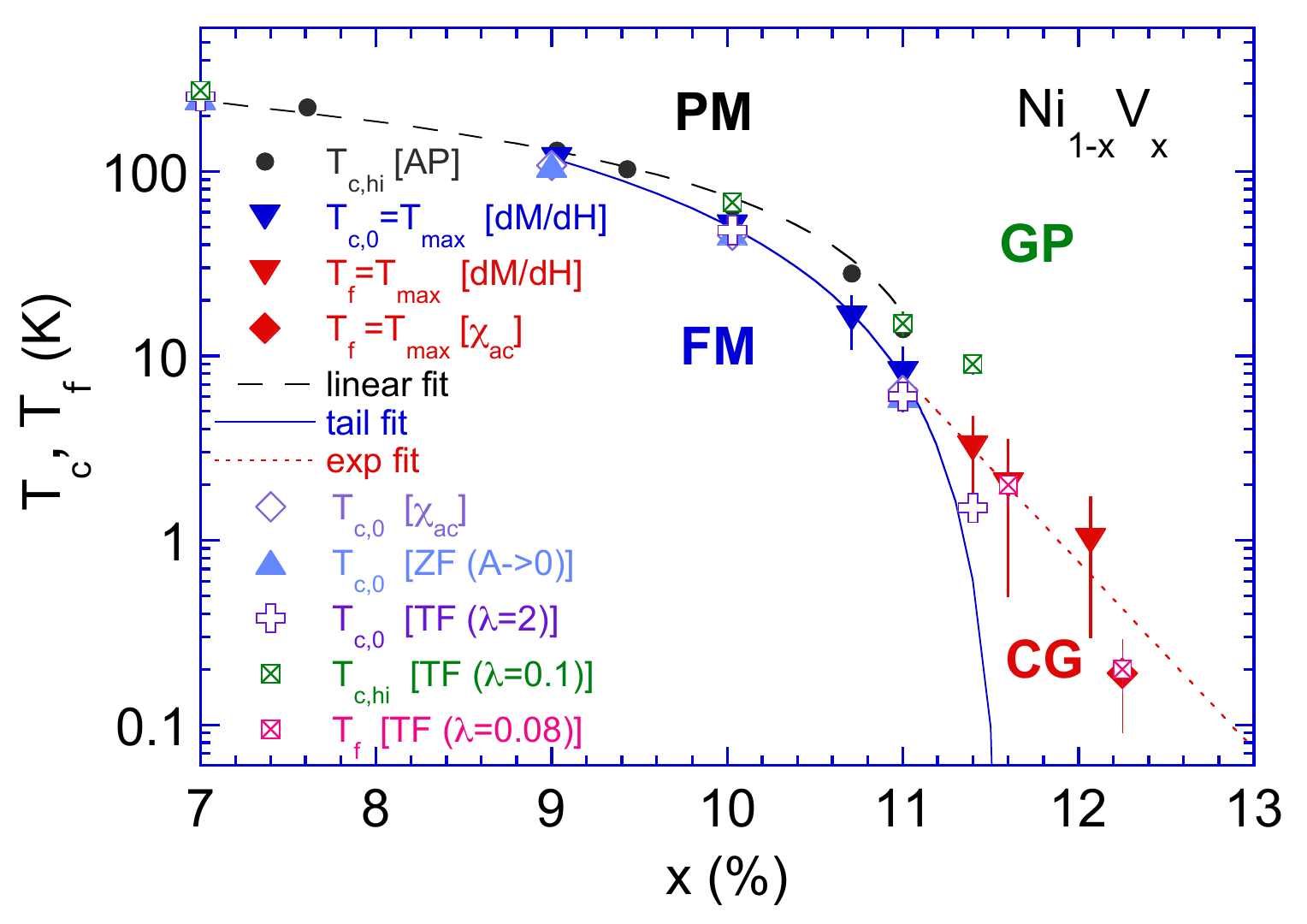}\hspace{1pc}%
\begin{minipage}[b]{19.5pc}\caption{\label{Fig5}Phase diagram of Ni$_{1-x}$V$_x$: Low-field ($T_{c,0}$) and high-field ($T_{c,hi}$) estimates of the phase boundary between paramagnet (PM) and ferromagnet (FM). Freezing temperature $T_f$ separating Griffiths phase (GP) and cluster glass (CG). Estimates from magnetization, susceptibility, and from ZF and TF $\mu$SR (for details see text). $T_{c,hi}(x)$ follows a linear fit (dashed line), $T_{f}(x)$ shows an exponential $x$-dependence (dotted line) while $T_{c,0}(x)$ close to the QPT follows  the ``tail fit'' discussed in the text (solid line). }
\end{minipage}
\end{figure}

\section{Summary}
The different phases and transitions observed in Ni$_{1-x}$V$_x$ are summarized in the phase diagram shown in Fig.\ 5.
The onset temperature $T_{c,0}$ of long-range FM order is determined from the maxima in $dM/dH(T,H)$ and $\chi_{ac}(T)$
as well as by the onset of strong muon depolarization in ZF and TF. The results from these methods agree well with each other.
In contrast, the values $T_{c,hi}$[AP] determined from high-field AP extrapolations work well for small dilutions $x$ but
 differ from $T_{c,0}$ as the QPT at $x_c$ is approached in this inhomogeneous system. This is caused by decoupled magnetic clusters that
align in strong fields but do not contribute to the spontaneous magnetization.
While $T_{c,hi}(x)$ initially shows a linear $x$-dependence, the low-field phase boundary $T_{c,0}(x)$ develops an exponential ``tail''
of the form $T_c(x)\sim \exp(-const (x-x_c)^{-\nu\psi})$ with $x_c=11.6~\%$ as proposed for disordered itinerant magnets \cite{vojta09}.

As was already shown in Ref.\ \cite{Ubaid10}, the behavior on the paramagnetic side of the transition, $x>x_c$,
is dominated by a distribution of magnetic clusters. In the Griffiths phase at higher temperatures, these clusters fluctuate independently.
Below a temperature $T_f$, they freeze into a short-range ordered cluster glass. $T_f$ decreases approximately exponentially with $x$.
Our present results provide strong indications that decoupled magnetic clusters also exist on the ferromagnetic side of the QPT,
roughly between $x=7\%$ and $x_c$. Evidence for such clusters comes from the reduced spontaneous moment at low $T$,
the absence of saturation in the $\lambda_{TF}$ data, and the multiple component description of the ZF $\mu$sR data \cite{muonpaper}.
This suggests an analog of the Griffiths phase on the ferromagnetic side of the QPT. A detailed quantitative analysis of this
phase will be performed elsewhere \cite{GPFM}.

\section{Acknowledgment}
We thank  W. Hayes for helpful discussions. This work was supported in part by OBR Research Challenge from KSU, by NSF under Grants No.\ DMR-0306766 and No.\ DMR-1205804, and EPSRC, UK. Part of this work was carried out at the Swiss Muon Source, Paul Scherrer Institute, CH and at ISIS, STFC Rutherford Appleton Laboratory, UK.

\section*{References}

\end{document}